\documentclass[12pt]{article} 
\usepackage{fullpage} 
\usepackage{setspace}
\usepackage{endnotes}
\usepackage{amssymb,amsfonts,amsmath}
\usepackage{graphicx}
\usepackage[position=bottom]{subfig}
\usepackage{booktabs} 
\usepackage{natbib}
\usepackage{rotating}
\usepackage[english]{babel}
\usepackage{hanging}

\begin{document} 

\setcounter{page}{1} 
\pagestyle{plain} 

\begin{center} 
\begin{Large} 
\textbf{How Untested Modeling Assumptions Influence the U.S. EPA's Estimates of Population-Level Ozone Exposure Risk}
\end{Large}

\vspace{1cm}

\large \textsc{Garrett Glasgow}\footnote{Director, NERA Economic Consulting, 4 Embarcadero Center, Suite 400, San Francisco, CA 94111, U.S.A. (garrett.glasgow@nera.com).}\\[0pt]
\vspace*{0.5cm} \textsc{Anne E. Smith}\footnote{Affiliated Consultant, NERA Economic Consulting, 2112 Pennsylvania Avenue NW, 4th Floor, Washington, DC 20037, U.S.A. (anne.smith.affiliate@nera.com).}\\[0pt]
\vspace*{0.5cm} April 6, 2025
\vspace*{1cm}

\large \textbf{ABSTRACT}

\end{center}

In recent reviews of the National Ambient Air Quality Standards (NAAQS) for ozone, the U.S. Environmental Protection Agency (U.S. EPA) has presented estimates of the health risks associated with ozone exposure.  One way in which the U.S. EPA calculates population-level ozone risk estimates is through a simulation model that calculates ozone exposures and the resulting lung function decrements for a simulated population.  This simulation model includes several random error terms to capture inter- and intra-individual variability in responsiveness to ozone exposure.  In this manuscript we undertake a sensitivity analysis examining the influence of untested assumptions about these error terms. We show that ad hoc bounds imposed on the error terms and the frequency of redrawing the intra-individual error terms have a strong influence on the population-level ozone exposure risk reported by the U.S. EPA.



\clearpage
\newpage

\section{Introduction}

In recent reviews of the National Ambient Air Quality Standards (NAAQS) for ozone, the U.S. Environmental Protection Agency (U.S. EPA) has presented estimates of the health risks associated with ozone exposure under alternative potential ozone standard levels [1-2].  A key short-term risk estimate considered by the U.S. EPA is based on a measure of lung function known as forced expiratory volume in one second (FEV$_{1}$), which is the volume of air that can be forcibly exhaled in one second after taking a full breath.  Decrements in FEV$_{1}$ (abbreviated dFEV$_{1}$) above a certain magnitude are thought to indicate an adverse health effect.

In past reviews the U.S. EPA has examined lung function risk across a variety of hypothetical air quality scenarios. These hypothetical air quality scenarios were created by adjusting actual measurements of ambient ozone in a variety of urban areas such that different standards for ozone would be just met – for example, a “75 parts per billion (ppb)” scenario sets hypothetical ambient ozone measurements to a level where a 75 ppb standard for ozone would be just met (the current NAAQS for ozone is 70 ppb, measured as the fourth-highest daily maximum 8-hour concentration averaged across three consecutive years).  For each hypothetical air quality standard, the U.S. EPA calculated population-level risk estimates as the estimated percentage of the population within various age groups (e.g., 5-18 years old) estimated to experience a dFEV$_{1}$ of a given magnitude (e.g., $\geq$ 10\%) at least a given number of times (e.g., on 2 different days) during an ozone season (e.g., March to November of a given year).  Reductions in these estimated percentages as hypothetical air quality improves are often cited by the U.S. EPA as evidence of health benefits that could be expected from tightening the ozone NAAQS.

One way in which the U.S. EPA calculates these risk estimates is through the Air Pollutants Exposure Model (APEX) [3-4].  APEX is a simulation model that calculates inhalation exposures associated with ambient pollutant concentrations for a simulated population of individuals.  Specifically, for each urban area and age group, APEX generates a set of simulated individuals that approximate a random sample of individuals living in the study area.  These simulated individuals experience different levels of ozone exposure based on their simulated activities and the simulated ozone scenario.  An exposure-response (E-R) function is then used to estimate dFEV$_{1}$ for each simulated individual at various points in time, and the simulated individual responses to ozone exposure are aggregated to produce population-level risk estimates (e.g., the percentage of individuals aged 5-18 who experienced dFEV$_{1} \geq$ 10\% at least once during the 2017 ozone season in Dallas).

The population-level risk estimates produced by the APEX model are the result of complicated individual-level simulations, with a variety of inputs and assumptions.  As with any such simulations, sensitivity analyses are crucially important to understanding how uncertainty around the inputs and modelling assumptions contributes to uncertainty in the outcome being modelled [5-6].  Some sensitivity analyses of the APEX model have been undertaken, primarily focused on uncertainties in inputs based on observable quantities, such as ozone exposure levels, the ventilation rate (the volume of gas entering or leaving the lungs over a given amount of time), the air-exchange rate (the number of times that the total air volume in a room or space is completely removed and replaced), and the selection of activities for simulated individuals [1-2, 7].

One aspect of the population-level risk estimates produced by the APEX model that is less well understood is the role of inter- and intra-individual variability in responsiveness to ozone exposure and in lung function generally. The ozone E-R functions used by APEX were derived from statistical models fit to controlled human exposure data [1-2, 8-9].  These E-R functions included several random terms to capture inter- and intra-individual variability in responsiveness to ozone exposure and lung function generally that were not captured by the systematic portion of the model.  These random error terms (which the U.S. EPA calls ``variability parameters'') are included in the APEX simulations as draws from mean zero normally distributed probability distributions.  Here we conduct sensitivity analyses focused on the influence of these error terms on the population-level risk estimates produced by the APEX model and reported by the U.S. EPA.

Specifically, the population-level risk estimates reported by the U.S. EPA are based on two largely unexamined assumptions about how the E-R function error terms should be implemented in the APEX model.  First, although the original E-R functions included in APEX did not assume any bounds on the mean zero normally distributed error terms [8-9], the APEX simulations bound all random draws from the error terms at $\pm 2$ standard deviations – random draws which fall outside these bounds were discarded, and new draws taken.  Second, although the E-R functions included in APEX were based on multiple (typically hourly) measurements per subject collected over 1 to 7.6 hours [10], the APEX simulations as implemented by the U.S. EPA redrew the error terms that control intra-individual variability in responsiveness to ozone exposure only once per simulated day.  These assumptions affect both the way in which changes in ozone exposure affect the population-level risk estimates and the baseline to which the population-level risk estimates should be compared (e.g., the percentage of individuals aged 5-18 who would experience dFEV$_{1} \geq$ 10\% at least once during the 2017 ozone season in Dallas even if there were no ambient ozone). Neither of these assumptions built into the APEX simulations is derived from observed data, and only limited sensitivity analyses have been undertaken in this area [1-2].

Our sensitivity analyses below focuses on (1) the influence of the ad hoc bounds imposed on the error terms in the APEX simulations, and (2) the influence of the frequency of redrawing the intra-individual error terms.  To keep our analyses tractable and simplify the presentation of the results, our sensitivity analyses focus on a single population-level ozone risk estimate presented by the U.S. EPA: the percentage of the population aged 5-18 estimated to experience dFEV$_{1} \geq$ 10\% for at least one day during the 2017 Dallas ozone season (March 1st through November 30th).  Focusing on this scenario is appropriate because NAAQS primary standards (the standards designed to protect public health) give extra consideration to the protection of ``sensitive'' groups such as children [2, 11], and because the most recent reported results for Dallas are comparable to those reported for other urban areas [2].  Similar results would be expected for different population-level risk estimates (e.g., experiencing dFEV$_{1} \geq$ 15\% four times during an ozone season) for different air quality scenarios, years, age groups, and urban areas.

\section{Methods}

\subsection{Preparing the APEX Simulations}

The latest version of the U.S. EPA’s Air Pollutants Exposure Model (APEX 5.24) was downloaded from the U.S. EPA’s website and installed on a Windows PC.  Although the APEX model is written in Fortran, it can be downloaded and run as a simple executable (.exe) file. APEX 5 is the same version of APEX used by the U.S. EPA in setting the most recent NAAQS standards.  These APEX results are presented in the U.S. EPA’s 2020 Policy Assessment (PA) [2].  The APEX download also includes most of the databases required to undertake an APEX simulation.

The replication files for the APEX simulations specific to the scenarios presented in the PA were obtained via email from the U.S. EPA. The PA presented APEX results averaged across 3 years (2015-2017) in 8 cities.  Only the files for the 75 ppb air quality scenario in Dallas in 2017 were used for the analyses below.

\subsection{APEX Calculation of Ozone Risk Estimates}

\subsubsection{Generating APEX ``Events''}

The APEX model generates detailed simulations of individual ozone exposures by first generating a set of simulated individuals using Census-derived probability distributions for demographic characteristics.  Each simulated individual then experiences a sequence of ``events'' generated from detailed survey data on activity patterns.  Each event is an activity that takes place at a specific time and in a specific location for a duration ranging from 1 to 60 minutes (e.g., a simulated individual might commute to work by car for 30 minutes, then work in an office for 60 minutes, and so on).

For each simulated individual, this sequence of events is combined with data on the hourly ozone concentrations relevant to the urban area and ozone scenario being studied.  The ambient ozone concentration and the level of exertion associated with each event, along with the simulated individual’s age, are then used as inputs into an exposure-response (E-R) function to estimate the lung function decrement (dFEV$_{1}$) for each simulated individual’s sequence of events.

\subsubsection{The APEX E-R Function}

The ozone E-R function used by APEX is derived from a statistical model fit to controlled human exposure data [8-9].  ``Exposure'' in this model is defined as the cumulative inhaled quantity of ozone over prior hours, with a decay in the effect of ozone that was inhaled at an increasingly distant point in time.   Following the notation in McDonnell et al. (2012) [9], the percent decrement in FEV$_{1}$ due to ozone exposure for an individual at time $t_{1}$ is calculated as:
\begin{multline}
dFEV_{1} = \exp{(U)} \Biggl[ \frac{\beta_{1} + \beta_{2}(A - \bar{A}) + \beta_{8}(BMI = \overline{BMI})}{1 + \beta_{4} \exp{(- \beta_{3} X_{t1})}} \\
 - \frac{\beta_{1} + \beta_{2}(A - \bar{A}) + \beta_{8}(BMI = \overline{BMI})}{1 + \beta_{4}} \Biggr] + E
\end{multline}
where $A$ is age in years, $BMI$ is body mass index, and $X(t_{1})$ is a measure of the accumulated ozone exposure at time $t_{1}$, which is calculated as:

\begin{equation}
X(t_{1}) = X(t_{0}) \exp{(- \beta_{5} (t_{1} - t_{0}))} +  \frac{C(t_{1})}{\beta_{5}} V(t_{1})^{\beta_{6}} [1 - \exp{(- \beta_{5} (t_{1} - t_{0}))]}
\end{equation}
where $C(t_{1})$ and $V(t_{1})$ are the constant exposure concentration and ventilation rates for the event from time $t_{0}$ to time $t_{1}$, and $V(t_{1})$ is measured as liters of air expired per minute normalized by body surface area. The term $X(t_{1})$ increases with concentration and ventilation rate, and allows for removal of ozone through a decay factor.  The accumulated ozone exposure $X(t_{1})$ that enters the calculation of dFEV$_{1}$ has an E-R threshold such that $X(t_{1})$ = max(0,  $X(t_{1}) - \beta_{9}$).  The parameter $\beta_{9}$ is the estimated E-R threshold; $X$ must exceed this value before dFEV$_{1}$ can exceed zero.

In the case of the APEX simulations, each event for a simulated individual is a time interval with a fixed exposure concentration and level of exertion, which allows for an analytic calculation of the dFEV$_{1}$ associated with each event.
 
The term contained in the square brackets in Eqn. 1 describes a sigmoidal E-R curve, with no expected response to ozone until an accumulated exposure threshold is crossed.  Differences across individuals in sensitivity to ozone are captured through the subject-specific random effect $U$, which is specified as a mean-zero normal distribution that varies across individuals, but not over time.  Multiplying the term in square brackets by the exponentiated value of $U$ for each individual flattens or steepens that individual’s E-R curve – individuals with greater sensitivity to ozone (a larger value of $U$) will have a steeper E-R curve.

All remaining variance in lung function that the estimated E-R function cannot predict is captured through the error term $E$. This error term is additive, increasing or decreasing dFEV$_{1}$ for each individual at each point in time.  It is possible for an individual to have a non-zero lung function decrement even if ozone exposure falls below the E-R threshold.  Two different specifications of the error term have been used in previous APEX risk calculations.  The first simply specifies $E$ as equal to $\nu1$, a mean-zero normal distribution that varies across individuals and time.  The E-R function with this specification is sometimes known as the ``MSS 2012'' E-R function, after McDonnell et al. (2012) [8].  The second specifies $E$ as equal to $\nu1 + \nu2 \times (\exp{(U) \times [M])}$,  where $M$ is the term contained in the square brackets in Eqn. 1 and both $\nu1$ and $\nu2$ are mean-zero normal distributions that vary across individuals and time.  The E-R function with this specification is sometimes known as the ``MSS 2013'' E-R function, after McDonnell et al. (2013) [9].  The MSS 2013 specification partitions the error term into two parts, one of which allows the intra-individual variance in lung function to increase in proportion to exposure and individual sensitivity to ozone and one of which reflects all other unexplained variance in lung function. The MSS 2013 specification results in a better fit to the controlled human exposure data used to estimate the E-R function than the MSS 2012 specification.

The model parameters $\beta$ and the variances of the error terms of $U$, $\nu1$, and $\nu2$ were taken from the ozone E-R functions as estimated on controlled human exposure data [8-9].  The standard deviations in the MSS 2012 E-R function were approximately 0.96 for $U$ and 4.13 for $\nu1$. while the standard deviations on the error terms in the MSS 2013 E-R function were approximately 1.06 for $U$, 3.02 for $\nu1$, and 1.47 for $\nu2$ [9]. The APEX simulations do not consider the statistical uncertainty in these estimated parameters and simply treat them as point estimates [12].  The APEX simulations used the estimated $\beta$s in Eqns. 1 and 2 above, and generated values of $U$, $\nu1$, and $\nu2$ by taking random draws from their estimated distributions. If bounds were placed on these distributions, any random draws outside the bounds were discarded and replaced with new draws.  The values of $U$ were generated by taking a single random draw for each simulated individual, while the values of $\nu1$ and $\nu2$ were generated by taking a new random draw for each simulated individual either each day or each hour of the simulated ozone season.

\subsubsection{Aggregating to Population-Level Risk}

The APEX simulations calculate dFEV$_{1}$ for each simulated individual’s sequence of events. These simulated individual responses to ozone exposure are then aggregated to produce various population-level ozone risk estimates (e.g., the percentage of individuals ages 5-18 that experienced dFEV$_{1} \geq$ 10\% at least once in an ozone season).

\subsection{Editing the APEX Input Files}

The APEX model inputs are contained in text files that can be user edited.  Two input text files were edited in the analyses presented below.

The first was the APEX Control Options file, which specifies the names of the input and output files, as well as a set of parameters specific to the APEX run.  The Control Options file for children (ages 5-18) under the 75 ppb air quality scenario in Dallas in 2017 was ``cof\_CSA206S75Y2017Child.txt.''  Edits to this file allow for changing the number of simulated individuals (line 59) and for specifying whether the draws of $\nu1$ and $\nu2$ in the ozone exposure-response function are drawn hourly or daily (lines 139-140).  The default values used by the U.S. EPA were 60,000 simulated individuals and new draws of $\nu1$ and $\nu2$ for each simulated day. Aside from these lines, and changing file names to create new output files and to read in the modified Physiological Parameters file (described below), all other Control Option inputs were unchanged from the file used to generate the results presented in the PA.

The second was the Physiological Parameters file, which contains tables of age- and gender-specific physiological parameters.  The Physiological Parameters file was ``Physiology051619\_Ufixed.txt.''  Edits to this file allow for changes in the bounds imposed on draws of $U$, $\nu1$, and $\nu2$ in the ozone E-R function (lines 1347-1352).  The default values were to bound the normally distributed draws of $U$, $\nu1$, and $\nu2$  at $\pm 2$ standard deviations. Sensitivity analyses in which ozone exposure has no effect on dFEV$_{1}$ were created by setting $\beta_{3}$ in the E-R function in Eqn. 1 to zero, such that ozone exposure has no effect on lung function (lines 1333-1334).  Aside from these lines, all other Physiological Parameters inputs were unchanged from the file used to generate the results presented in the PA.

\section{Results}

\subsection{Removing the Bounds on the E-R Function Error Terms}

Table 1 reveals that the ad hoc bounds on the APEX E-R function error terms have a clear effect on the population-level ozone risk estimates calculated by the U.S. EPA.

\begin{table*}[h]
\centering
\captionsetup{labelfont={bf}}
\caption{Population-Level Ozone Risk Estimates Generated by APEX Simulations Under Different Error Term Bound Assumptions and Frequency of Error Term Draws}
\begin{tabular}{lcc}
\toprule
Simulation Scenario						& Daily Draws 	& Hourly Draws	\\
\midrule
MSS 2012 E-R, 75 ppb Scenario 		& 				&				\\	
 \ \ \ $\pm 2$ Standard Deviations 	& 13.72		& 18.59		\\	
 \ \ \ Unbounded						& 91.44		& 100.00		\\
MSS 2013 E-R, 75 ppb Scenario 		& 				&				\\	
 \ \ \ $\pm 2$ Standard Deviations 	& 16.94		& 22.68		\\	
 \ \ \ Unbounded						& 37.47		& 96.30		\\	
MSS 2012 E-R, 0 ppb Scenario 		& 				&				\\	
 \ \ \ $\pm 2$ Standard Deviations 	& 0.00		& 0.00		\\	
 \ \ \ Unbounded						& 88.11		& 100.00		\\
MSS 2013 E-R, 0 ppb Scenario 		& 				&				\\	
 \ \ \ $\pm 2$ Standard Deviations 	& 0.00		& 0.00		\\	
 \ \ \ Unbounded						& 12.11		& 95.24		\\
\bottomrule
\end{tabular}
\begin{flushleft} \emph{\textbf{Notes:} Numbers are the percentage of the simulated population ages 5-18 experiencing dFEV$_{1} \geq$ 10\% at least once during the Dallas 2017 ozone season. All simulations used 60,000 simulated individuals.} 
\end{flushleft}
\end{table*}

Using the MSS 2012 E-R function and setting bounds of $\pm 2$ standard deviations on both error terms ($U$ and $\nu1$), the APEX simulation predicts 13.72\% of individuals ages 5-18 would experience dFEV$_{1}$ greater than 10\% for at least one day during the 2017 Dallas ozone season under the 75 ppb air quality scenario when taking daily draws from the error terms.  This percentage increases to 18.59\% when taking hourly draws from the error terms.  Using the MSS 2013 E-R function and setting bounds of $\pm 2$ standard deviations on all three error terms ($U$, $\nu1$, and $\nu2$), the APEX simulation predicts 16.94\% of individuals ages 5-18 would experience dFEV$_{1}$ greater than 10\% for at least one day during the 2017 Dallas ozone season under the 75 ppb air quality scenario when taking daily draws from the error terms.  This percentage increases to 22.68\% when taking hourly draws from the error terms.  The increase when using hourly draws is expected, as taking hourly draws increases the probability of a draw that creates a dFEV$_{1}$ greater than 10\%.

To test the effect of the ad hoc bounds on the error terms in the APEX E-R function, we re-estimated the scenarios described above removing the bounds on the error terms.  As a further test, we also estimated simulations under scenarios in which there was no ozone.  The zero-ozone scenario is the baseline to which the population-level ozone risk estimates should be compared.

Under the 75 ppb air quality scenario for the MSS 2012 E-R function, removing the bounds on the error terms increases the population-level risk estimate by nearly a factor of 7 (from 13.72\% to 91.44\%) when taking daily draws, and the risk estimate reaches 100\% when taking hourly draws. Under the 75 ppb air quality scenario for the MSS 2013 E-R function, removing the bounds on the error terms more than doubles the population-level risk estimate (from 16.94\% to 37.47\%) when taking daily draws, and more than quadruples it (from 22.68\% to 96.30\%) when taking hourly draws.

Under the zero-ozone scenario the APEX simulation does not produce lung function decrements with the bounds on the error terms in place.  This is because the error terms $U$ and $\nu2$ must interact with ozone exposure to have an effect, and bounds of $\pm 2$ standard deviations on $\nu1$ restrict dFEV$_{1}$ to a maximum of $\pm 8.26\%$ in the MSS 2012 E-R function and $\pm 6.04\%$ for the MSS 2013 E-R function without ozone exposure.  Removing the bounds on $\nu1$ allows for dFEV$_{1}$ decrements greater than 10\% even without ozone exposure.  With daily draws from the error terms and no bounds, the population-level risk estimate for the MSS 2012 E-R function under the zero-ozone scenario was 88.11\%, and with hourly draws the population-level risk estimate was 100\%.  With daily draws from the error terms and no bounds, the population-level risk estimate for the MSS 2013 E-R function under the zero-ozone scenario was 12.11\%, and with hourly draws the population-level risk estimate was 95.24\%.  

The results from the zero-ozone scenario demonstrate that the bounds imposed by the APEX model affect not only the population-level risk estimate for a given ozone exposure, but also the baseline to which the population-level risk estimates should be compared. This baseline becomes non-zero once the error term bounds are removed, changing the incremental risk due to ozone exposure. For example, with hourly draws from the error terms and no bounds, the population-level risk estimate for the MSS 2013 E-R function only increases by 1.06 percentage points (96.30\% - 95.24\%) when moving from the zero-ozone scenario to the 75 ppb scenario.

Table 1 also demonstrates that the frequency of the draws of $\nu1$ and $\nu2$ are important, with more frequent draws producing higher population-level risk estimates.  For a given set of bounds, more frequent draws from the error terms will be more likely to result in at least one draw that produces a decrement above a given threshold.  The increase in the risk estimates as repeated draws are taken from the error terms can be demonstrated analytically.  For simplicity, consider the zero-ozone scenario, where decrements can only be produced through the draw of $\nu1$.  In the APEX simulations as implemented by the U.S. EPA $\nu1$ is a truncated normal distribution with mean zero and bounds at $\pm b$ standard deviations.  The probability that one draw from this distribution will not produce a decrement that is $x$ standard deviations or larger (assuming that $b > x$) is given by:

\begin{equation}
P_{nd} = \frac{\Phi(x) - \Phi(-b)}{\Phi(b) - \Phi(-b)}
\end{equation}
The probability of at least one decrement $\geq x$ over $t$ draws is then given by:

\begin{equation}
P_{d} = 1 - P_{nd}^{t}
\end{equation}
In a large population, $P_{d}$ will equal the fraction of the simulated population experiencing at least one decrement of size $x$.  Since $b > x$, $P_{nd}$ is always less than 1, and thus the risk estimates are increasing in the number of draws $t$.

\subsection{Varying Bounds from $\pm 0$ to $\pm 4$ Standard Deviations on the E-R Function Error Terms}

The true amount of variance in lung function over time is certainly subject to physiological bounds (at the minimum, lung function decrements cannot be larger than 100\%).  However, the way in which physiological bounds should be translated into bounds on the error terms in the APEX E-R functions is unknown.  Thus, we examine the effect of a range of potential bounds on the population-level ozone risk estimates.

The APEX simulations as implemented by the U.S. EPA sets bounds of $\pm 2$ standard deviations on all error terms in the E-R functions.  We consider the effect on the population-level ozone risk estimated when varying the bounds on each error term from $\pm 0$ (meaning that error term has no effect, since it is mean zero) to $\pm 4$ standard deviations (which for a normal distribution will produce results nearly equivalent to those from an unbounded distribution), while holding the bounds on the other error terms at $\pm 2$ standard deviations.  We used 10,000 simulated individuals for each simulation (the U.S. EPA has also used 10,000 simulated individuals for some of its sensitivity analyses [2]), and considered both daily and hourly draws from the error terms $\nu1$ and $\nu2$.

\subsubsection{Individual Ozone Sensitivity ($U$)}

The error term $U$ allows for differences in individual sensitivity to ozone.  Figure 1 presents the change in the population-level ozone risk estimate as the bounds on $U$ vary from $\pm 0$ to $\pm 4$ standard deviations for the MSS 2012 E-R function, while Figure 2 presents the same change for the MSS 2013 E-R function.  Note that $U$ is only drawn once for each simulated individual in APEX – the ``daily'' and ``hourly'' draws refer to the draws of $\nu1$ and $\nu2$.

\begin{figure*}
\centering
\captionsetup{labelfont={bf}}
    \caption{Population-Level Ozone Risk Estimates Generated by APEX Simulations as Bounds on $U$ Vary}
    \centering
    \subfloat[MSS 2012 E-R]{{\includegraphics[width=0.5\textwidth]{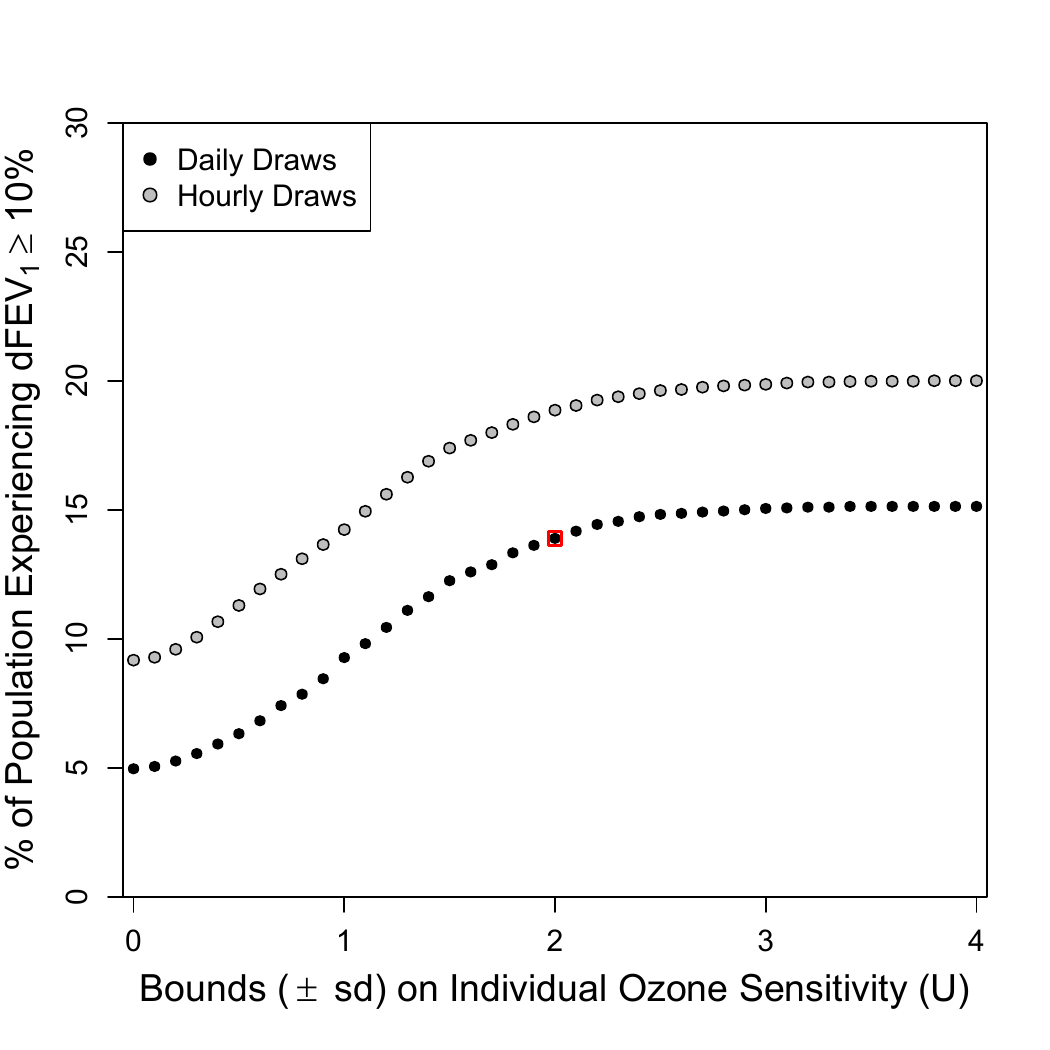} }}
    \qquad
    \subfloat[MSS 2013 E-R]{{\includegraphics[width=0.5\textwidth]{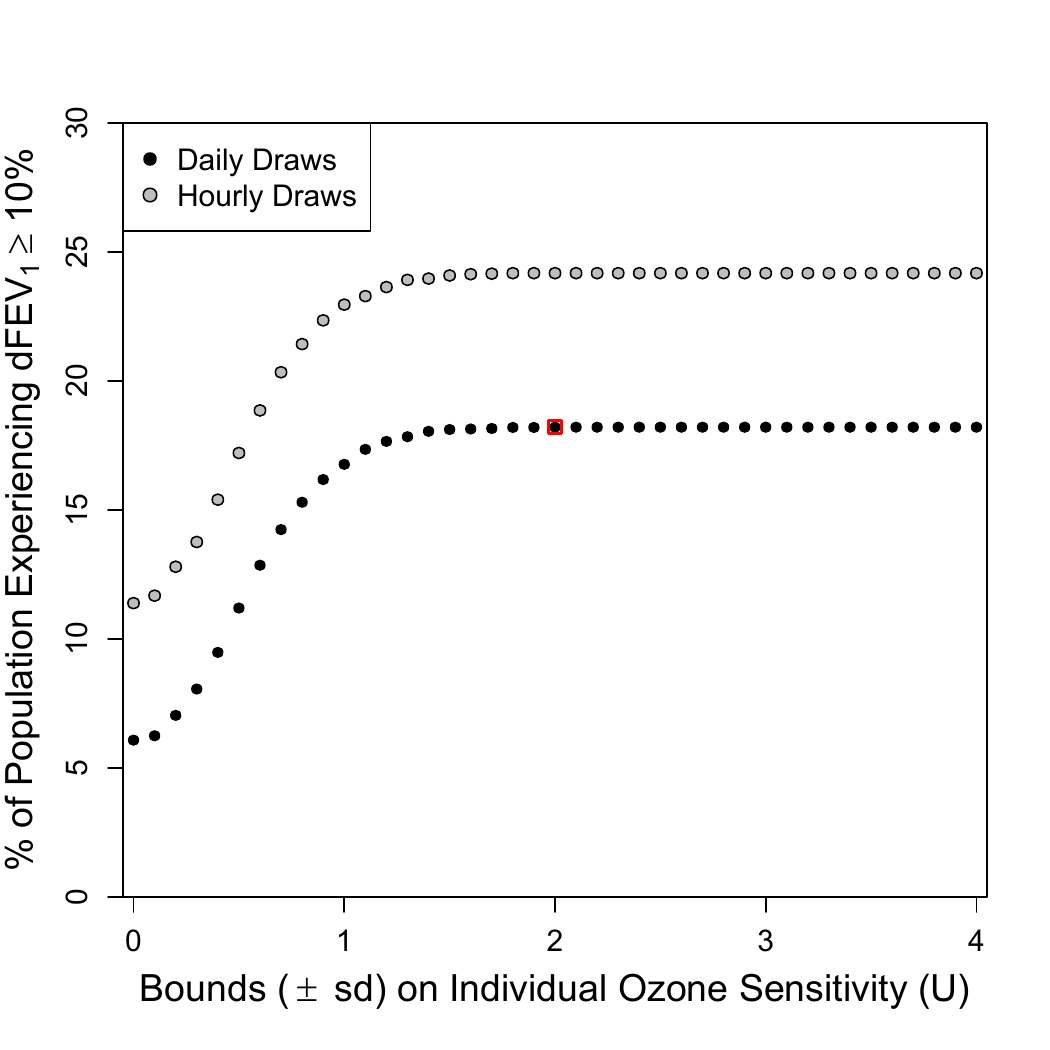} }}
 \\
\begin{flushleft} \emph{\textbf{Notes:} ``sd'' = standard deviations. Points indicate the percentage of the simulated population ages 5-18 experiencing dFEV$_{1} \geq$ 10\% at least once during the Dallas 2017 ozone season. The red square indicates the result of the assumptions used by the U.S. EPA in its APEX simulations.  All simulations used 10,000 simulated individuals.}
\end{flushleft}
\end{figure*}

Figure 1 shows that for both the MSS 2012 and MSS 2013 E-R functions, as the bounds on $U$ increase the population-level risk estimate increases, but the effect plateaus soon after $\pm 1$ standard deviations.  Across the range of bounds considered here, Figure 1(a) shows that for the MSS 2012 E-R function, the population-level risk estimates increased from 4.97\% to 15.14\% (by 10.17 percentage points) for daily draws of $\nu1$ and from 9.18\% to 20.01\% (by 10.83 percentage points) for hourly draws of $\nu1$.  Figure 1(b) shows that for the MSS 2013 E-R function, the equivalent results are an increase from 6.08\% to 18.21\% (by 12.13 percentage points) for daily draws of $\nu1$ and $\nu2$ and from 11.39\% to 24.18\% (by 12.79 percentage points) for hourly draws of $\nu1$ and $\nu2$.

\subsubsection{Time Varying Ozone Sensitivity ($\nu2$)}

The error term $\nu2$ allows for intra-individual variance in lung function based on exposure and individual sensitivity to ozone.  Figure 2 presents the change in the population-level ozone risk estimate as the bounds on $\nu2$ vary from $\pm 0$ to $\pm 4$ standard deviations.  Greater variance in the draws of $\nu2$ will produce greater intra-individual variance in lung function as ozone exposure increases. 

\begin{figure*}[h]
\centering
\captionsetup{labelfont={bf}}
    \caption{Population-Level Ozone Risk Estimates Generated by APEX Simulations as Bounds on $\nu2$ Vary}
    \centering
    \includegraphics[width=0.5\textwidth]{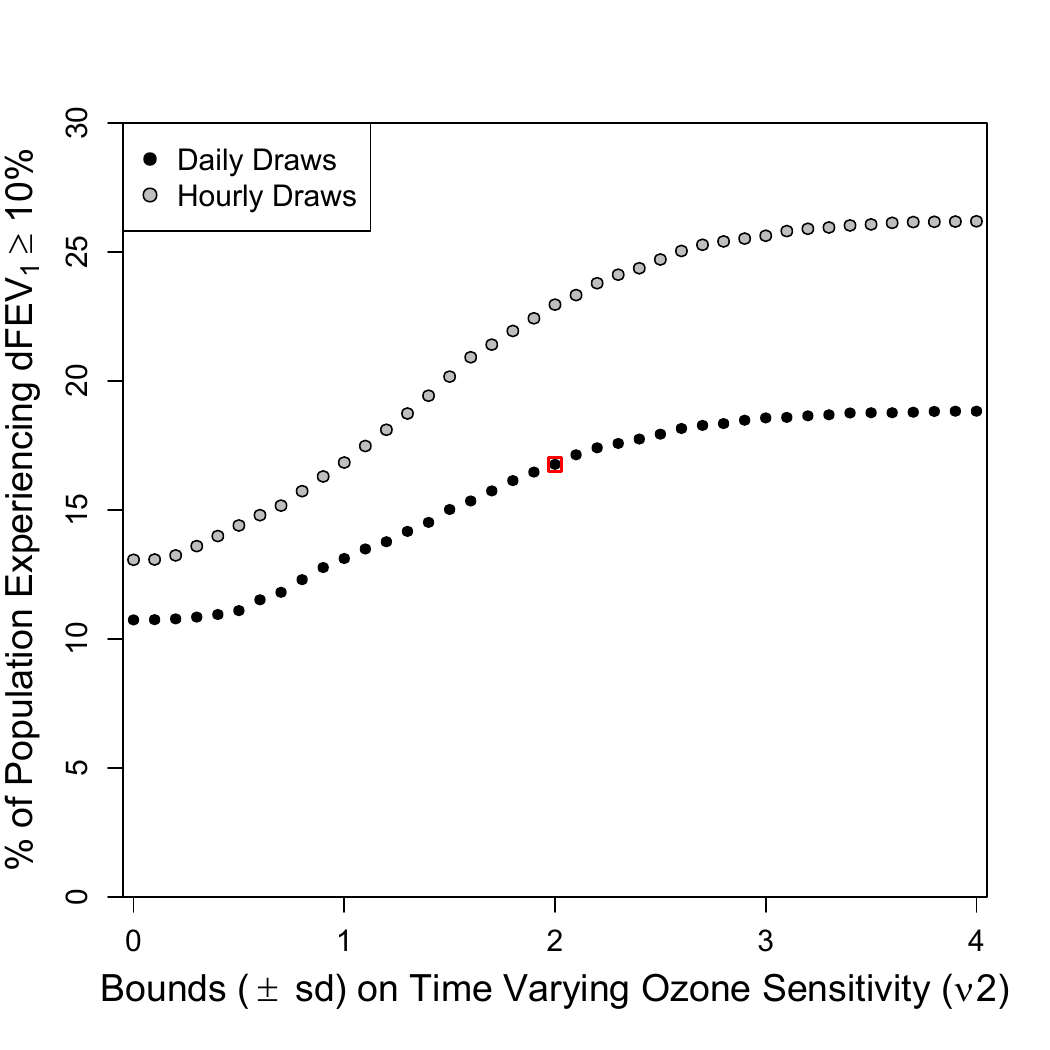}
 \\
\begin{flushleft} \emph{\textbf{Notes:} ``sd'' = standard deviations. Points indicate the percentage of the simulated population ages 5-18 experiencing dFEV$_{1} \geq$ 10\% at least once during the Dallas 2017 ozone season. The red square indicates the result of the assumptions used by the U.S. EPA in its APEX simulations. All simulations used 10,000 simulated individuals and the MSS 2013 E-R function.} 
\end{flushleft}
\end{figure*}

As with the error term $U$, as the bounds on $\nu2$ increase the population-level risk estimate increases, but the effect plateaus, in this case at approximately $\pm 3$ standard deviations.  Across the range of bounds considered here, the population-level risk estimates increased from 10.74\% to 18.83\% (by 8.09 percentage points) for daily draws of $\nu1$ and $\nu2$, and from 13.07\% to 26.19\% (by 13.12 percentage points) for hourly draws of $\nu1$ and $\nu2$. There are no equivalent results for the MSS 2012 E-R function, as error term $\nu2$ only appears in the MSS 2013 E-R function.  

\subsubsection{Time Varying Lung Function ($\nu1$)}

The error term $\nu1$ allows for greater intra-individual variance in lung function regardless of ozone exposure.  Figure 3 presents the change in the population-level ozone risk estimate as the bounds on $\nu1$ vary from $\pm 0$ to $\pm 4$ standard deviations.

\begin{figure*}
\centering
\captionsetup{labelfont={bf}}
    \caption{Population-Level Ozone Risk Estimates Generated by APEX Simulations as Bounds on $\nu1$ Vary}
    \centering
    \subfloat[MSS 2012 E-R]{{\includegraphics[width=0.5\textwidth]{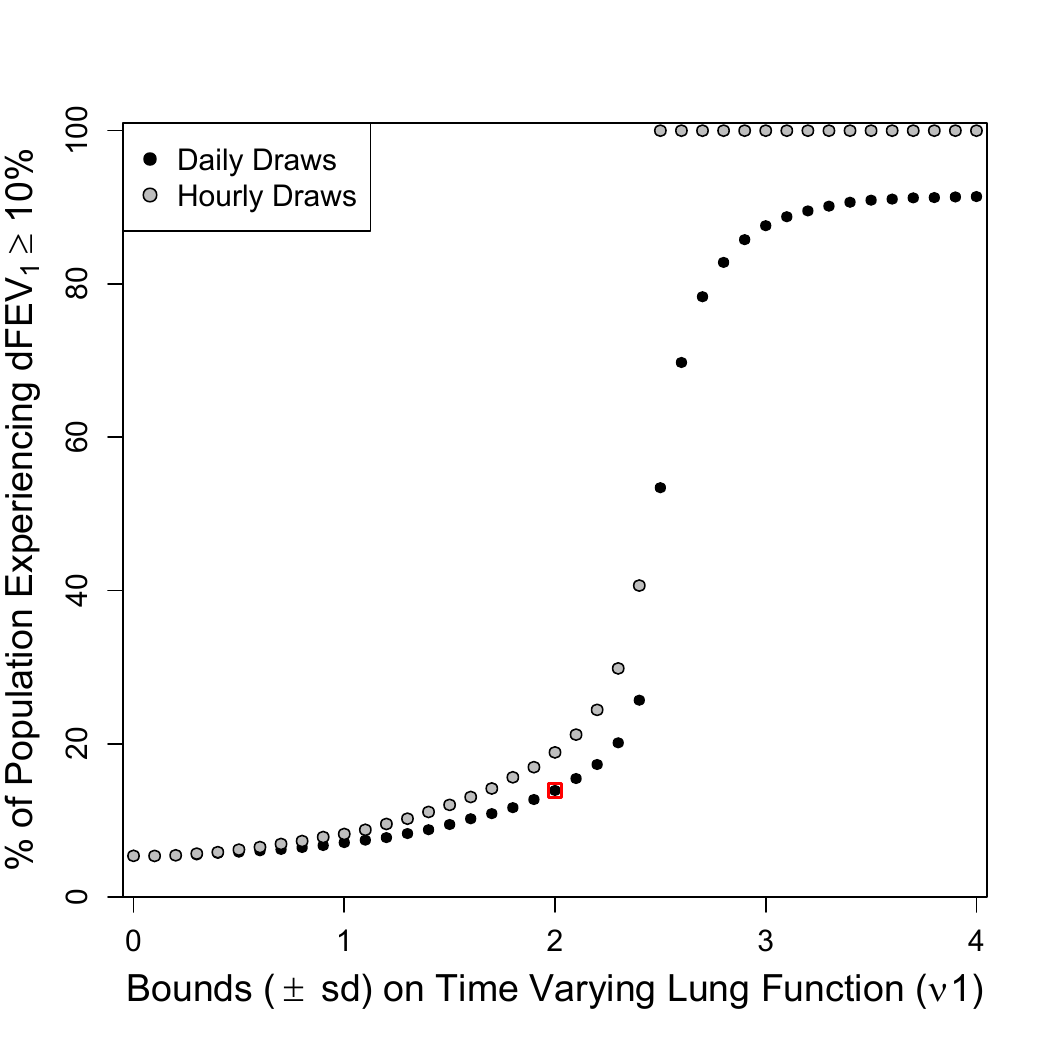} }}
    \qquad
    \subfloat[MSS 2013 E-R]{{\includegraphics[width=0.5\textwidth]{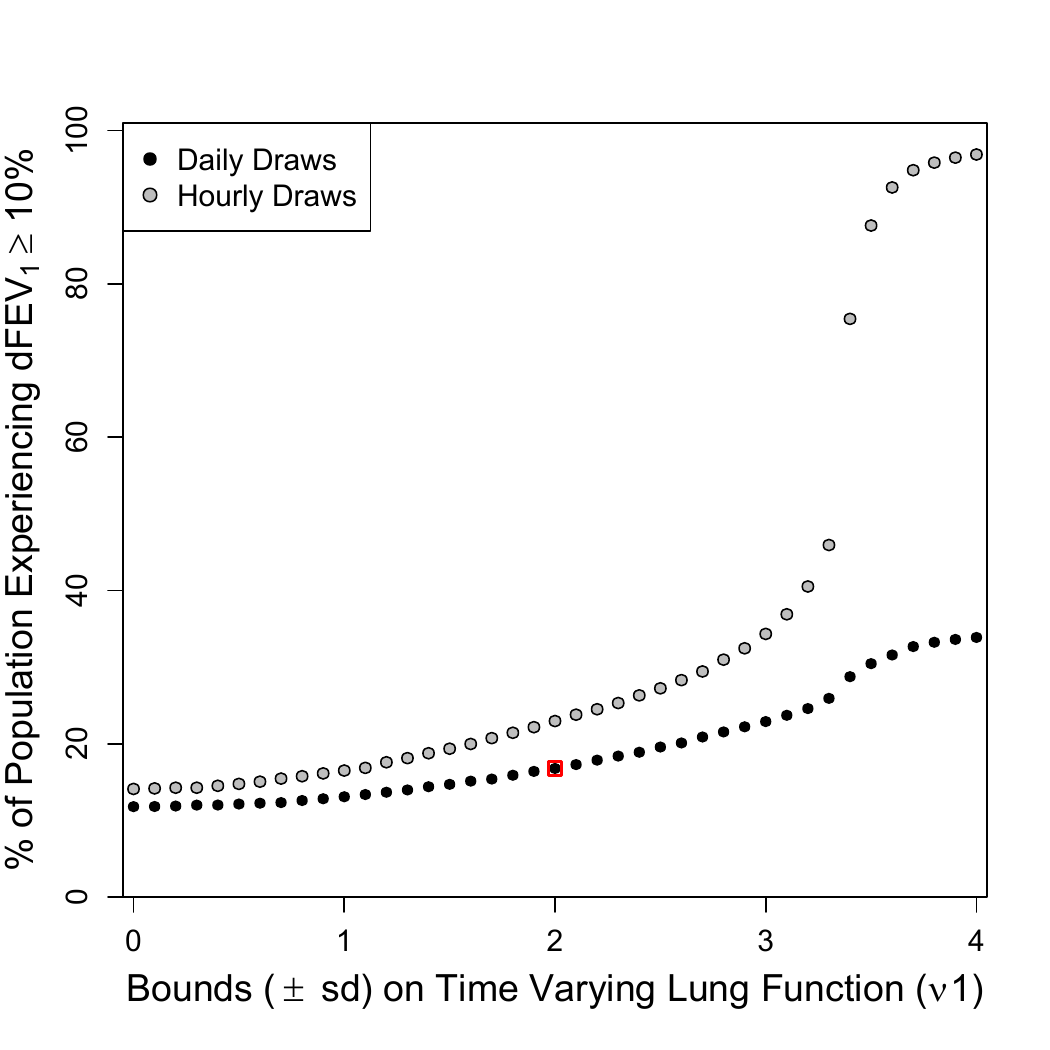} }}
 \\
\begin{flushleft} \emph{\textbf{Notes:} ``sd'' = standard deviations. Points indicate the percentage of the simulated population ages 5-18 experiencing dFEV$_{1} \geq$ 10\% at least once during the Dallas 2017 ozone season. The red square indicates the result of the assumptions used by the U.S. EPA in its APEX simulations.  All simulations used 10,000 simulated individuals.}
\end{flushleft}
\end{figure*}

As expected, the population-level risk estimates increase as the bounds on $\nu1$ widen.  Figure 3(a) shows that for the MSS 2012 E-R function, the population-level risk estimates increased from 5.37\% to 91.40\% (by 86.03 percentage points) for daily draws of $\nu1$ as the bounds widened from $\pm 0$ to $\pm 4$ standard deviations, and from 5.37\% to 100\% (by 94.63 percentage points) for hourly draws of $\nu1$ across the same range. Figure 3(b) shows that for the MSS 2013 E-R function, the population-level risk estimates increased from 11.79\% to 33.87\% (by 22.08 percentage points) for daily draws of $\nu1$ and $\nu2$ as the bounds widened from $\pm 0$ to $\pm 4$ standard deviations, and from 14.10\% to 96.89\% (by 82.79 percentage points) for hourly draws of $\nu1$ and $\nu2$ across the same range. Note that unlike the MSS 2012 E-R function, the population-level risk estimate for bounds of $\pm 0$ standard deviations on $\nu1$ in the MSS 2013 E-R function varies between daily and hourly draws due to the draws of $\nu2$. 

For the MSS 2012 E-R function, there is a clear increase in the risk estimate beyond bounds of approximately $\pm 2.5$ because at this point dFEV$_{1}$ decrements greater than 10\% can be created by $\nu1$ alone, regardless of ozone exposure (the standard deviation on $\nu1$ in the MSS 2012 E-R function is approximately 4.13, and 4.13 $\times$ 2.5 $\approx$ 10.3).  Similarly, for the MSS 2013 E-R function there is a clear increase in the risk estimate beyond bounds of approximately $\pm 3.3$ (the standard deviation on $\nu1$ in the MSS 2013 E-R function is approximately 3.02, and 3.02 $\times$ 3.3 $\approx$ 10).  This effect is more noticeable with hourly draws than daily draws, as with hourly draws there are 24 times as many chances to take a draw of $\nu1$ that results in a decrement. Note that when lung function decrements can occur without ozone exposure, the incremental risk due to ozone exposure will be lower than the risk estimate reported by APEX, which only reports the number of simulated individuals experiencing a decrement regardless of the cause. 

\clearpage
\newpage

\subsection{Combinations of Bounds on $\nu1$ and $\nu2$}

The APEX simulations as implemented by the U.S. EPA sets bounds of $\pm 2$ standard deviations on all error terms.  We consider the effect on the population-level ozone risk estimated through the MSS 2013 E-R function when setting various bounds on the draws of $\nu1$ and $\nu2$ simultaneously, while holding the bounds on the distribution of $U$ constant at $\pm 2$ standard deviations (since Section 3.2.1 above shows that the risk estimates are similar across a wide range of bounds for $U$, we do not consider it further below).  There are no equivalent results for the MSS 2012 E-R function, as the error term $\nu2$ only appears in the MSS 2013 E-R function. 

Table 2 presents the population-level ozone risk estimates for combinations of bounds for $\nu1$ and $\nu2$.

\begin{table*}[h]
\centering
\captionsetup{labelfont={bf}}
\caption{Population-Level Ozone Risk Estimates Generated by APEX Simulations Under Different Error Term Bound Assumptions and Frequency of Error Term Draws}
\begin{tabular}{l|ccccc}
\toprule
Daily:			& $\nu1 \pm 0$ sd	& $\nu1 \pm 1$ sd	& $\nu1 \pm 2$ sd 	& $\nu1 \pm 3$ sd  & $\nu1 \pm 4$ sd \\
\midrule
$\nu2 \pm 0$ sd	& 7.17	& 8.05	& 10.43	& 13.96	& 26.16\\ 
$\nu2 \pm 1$ sd	& 8.83	& 10.02	& 12.98	& 17.48	& 29.47 \\
$\nu2 \pm 2$ sd	& 11.42	& 12.95	& \textbf{16.94} & 22.77 & 34.17 \\
$\nu2 \pm 3$ sd	& 12.62	& 14.28	& 18.76	& 24.76	& 35.88 \\
$\nu2 \pm 4$ sd	& 12.82	& 14.63	& 19.01	& 25.07	& 36.14 \\
\midrule
\midrule
Hourly:			& $\nu1 \pm 0$ sd	& $\nu1 \pm 1$ sd	& $\nu1 \pm 2$ sd 	& $\nu1 \pm 3$ sd  & $\nu1 \pm 4$ sd \\
\midrule
$\nu2 \pm 0$ sd	& 7.17	& 8.88	& 12.60	& 19.42	& 96.03 \\ 
$\nu2 \pm 1$ sd	& 9.95	& 11.88	& 16.63	& 26.39	& 96.36 \\
$\nu2 \pm 2$ sd	& 13.84	& 16.28	& 22.68	& 34.85	& 96.79 \\
$\nu2 \pm 3$ sd	& 15.79	& 18.60	& 25.62	& 38.14	& 96.94 \\
$\nu2 \pm 4$ sd	& 16.18	& 19.02	& 26.11	& 38.63	& 96.96 \\
\bottomrule
\end{tabular}
\begin{flushleft} \emph{\textbf{Notes:} ``sd'' = standard deviations. Numbers are the percentage of the simulated population ages 5-18 experiencing dFEV$_{1} \geq$ 10\% at least once during the Dallas 2017 ozone season. The bold face entry indicates the result of the assumptions used by the U.S. EPA in its APEX simulations.  All simulations used 60,000 simulated individuals and the MSS 2013 E-R function.} 
\end{flushleft}
\end{table*}

\section{Discussion}

Our sensitivity analyses show that the U.S. EPA's population-level ozone risk estimates are strongly influenced by the ad hoc bounds placed on the mean zero normally distributed error terms in the APEX simulations as implemented by the U.S. EPA, particularly when the bounds on the additive error term that allows for intra-individual variance in lung function regardless of ozone exposure ($\nu1$) become wide enough to allow for lung function decrements greater than a threshold of interest (e.g., 10\%) without ozone exposure.  As these bounds widen, the population-level risk estimate we consider here (dFEV$_{1} \geq$ 10\%) approaches 100\% of the population, which changes the baseline to which we should compare the population-level ozone risk estimates and thus the incremental risk due to ozone exposure.  Even across ``reasonable'' bounds on the error terms the population-level ozone risk estimates can vary widely. For example, Table 2 demonstrates that bounding the intra-individual error terms at $\pm 1$ standard deviation and taking daily draws of these error terms leads to a population-level risk estimate of 10.02\%, while bounding the intra-individual error terms at $\pm 3$ standard deviations and taking hourly draws of these error terms leads to a population-level risk estimate that is nearly four times higher (38.14\%).  The bounds imposed on the error terms in the U.S. EPA’s APEX simulations are not derived from observed data, and thus the ``true'' location of any such bounds on the error terms is unknown.  However, previous work on lung function suggests that inter- and intra-individual variability in dFEV$_{1}$ may be higher than the default APEX bounds allow [8, 13-15]. For example, variation in FEV$_{1}$ of up to 12\% between doctor’s visits is considered ``normal'' by some diagnostic standards [14-15].

Our sensitivity analyses also show that the frequency with which the intra-individual error terms are redrawn has a strong influence on the U.S. EPA's population-level ozone risk estimates. Figures 1-3 demonstrate that for any bounds on the error terms, more frequent draws of the intra-individual error terms lead to larger population-level risk estimates, and these risk estimates approach 100\% once the bounds on $\nu1$ become wide enough to allow for lung function decrements greater than a threshold of interest without ozone exposure.  As with the error term bounds, the frequency of redrawing the error terms is not derived from observed data, and the ``true'' frequency is unknown.  However, previous work on lung function demonstrates that there can be significant intra-individual variability in dFEV$_{1}$ within a single day [8, 13-15].  Further, the E-R functions included in APEX were estimated on data consisting of multiple measurements per subject collected over 1 to 7.6 hours [10], suggesting that at least some of the unexplained variability in dFEV$_{1}$ captured in the error terms consists of intra-individual variation within a single day.

A previous sensitivity analysis of the APEX model demonstrated that a small number of the randomly generated inputs based on observed data were responsible for most of the variation in simulated ozone exposure and dose [7].  The sensitivity analyses in our paper identify another important source of uncertainty, demonstrating that ad hoc assumptions about inter- and intra-individual variability in responsiveness to ozone exposure can have a significant effect on the population-level ozone risk estimates produced by the APEX model and reported by the U.S. EPA.  Neither assumption about the E-R function error terms imposed by the U.S. EPA in the APEX model (the error term bounds and the frequency with which to draw the error terms) is based on empirical data – this is an obvious area for future research and improvement of both the APEX simulations and the U.S. EPA's measures of population-level ozone exposure risk.

\clearpage
\newpage

\section*{References}

\leftskip 0.4in
\parindent -0.4in

\textbf{}

\textbf{[1]} U.S. EPA. Health risk and exposure assessment for ozone (final report). Research Triangle Park, NC: EPA Office of Air Quality Planning and Standards, EPA–452/R–14–004a. 2014.

\textbf{[2]} U.S. EPA. Policy assessment for the reconsideration of the ozone national ambient air quality standards. Research Triangle Park, NC: EPA Office of Air Quality Planning and Standards, EPA–452/R–20–001. 2020.

\textbf{[3]}	U.S. EPA. Air pollutants exposure model documentation (APEX, Version 5.2) volume I: User’s guide. EPA–452/R–19–005a. 2019a.

\textbf{[4]}	U.S. EPA. Air pollutants exposure model documentation (APEX, Version 5.2) volume II: Technical support document. EPA–452/R–17–001b. 2019b.

\textbf{[5]}	Razavi S, Jakeman A, Saltelli A, Prieur C, Iooss B, Borgonovo E, et al. 2021. The future of sensitivity analysis: An essential discipline for systems modeling and policy support. Environ Model Softw. 2021; 137:104954. 10.1016/j.envsoft.2020.104954 

\textbf{[6]} Nsoesie EO, Beckman RJ, Marathe MV. Sensitivity analysis of an individual-based model for simulation of influenza epidemics. PLoS ONE. 2012; 7(10): e45414. \\ doi:10.1371/journal.pone.0045414 

\textbf{[7]} Langstaff J, Glen G, Holder C, Graham S, Isaacs K. A sensitivity analysis of a human exposure model using the Sobol method.  Stoch Environ Res Risk Assess. 2022; 36: 3945–3960. 

\textbf{[8]} McDonnell WF, Stewart PW, Smith MV, Kim CS, Schelegle ES. Prediction of lung function response for populations exposed to a wide range of ozone conditions. Inhal. Toxicol. 2012; 24: 619–633.

\textbf{[9]} McDonnell WF, Stewart PW, Smith MV. Ozone exposure-response model for lung function changes: An alternate variability source. Inhal. Toxicol. 2013; 25: 348–353.

\textbf{[10]} McDonnell WF, Stewart PW, Smith MV. The temporal dynamics of ozone-induced FEV1 changes in humans: An exposure-response model. Inhal. Toxicol. 2007; 19: 483–494.

\textbf{[11]} Clean Air Act, 42 U.S. Code § 7408–7409.

\textbf{[12]} Glasgow G, Smith A. Uncertainty in the estimated risk of lung function decrements owing to ozone exposure. J Expo Sci Environ Epidemiol. 2017; 27: 535–538.

\textbf{[13]} Becklake MR. Concepts of normality applied to the measurement of lung function. Am J Med 1986; 80: 1158–1164.

\textbf{[14]} Dean BW, Birnie EE, Whitmore GA, Vandemheen KL, Boulet LP, FitzGerald JM, et al. Between-visit variability in FEV1 as a diagnostic test for asthma in adults. Ann Am Thorac Soc. 2018; 15:1039–1046.

\textbf{[15]} Pellegrino R, Viegi G, Brusasco V, Crapo RO, Burgos F, Casaburi R, et al. Interpretative strategies for lung function tests.  Eur Respir J 2005; 26: 948–968.

\end{document}